\documentstyle[prl,aps,multicol,epsfig]{revtex}

\begin{document}

\def\bb#1{{\bf #1}}
\def\pd#1#2{{\partial #1\over\partial #2}}

\title{\bf  Evolution of  rarefaction pulses into  vortex rings. }
\author{Natalia G. Berloff\\
Department of Mathematics\\
University of California, Los Angeles, CA, 90095-1555\\ 
{\it nberloff@math.ucla.edu}}
\date {\today}
\maketitle
\begin {abstract}

The 
two-dimensional solitary waves of the Gross-Pitaevskii  equation in 
the  Kadomtsev-Petviashvili  limit   
are unstable with respect to three-dimensional perturbations. 
We elucidate the stages in the evolution of such solutions subject 
to   perturbations perpendicular to the direction of  motion. 
Depending on the energy (momentum) and the wavelength of the perturbation  
different types of  three-dimensional solutions  emerge. In particular, 
we present   new  periodic solutions having very small  energy and 
momentum per period. These solutions also  become  unstable and 
this secondary instability leads to  vortex ring nucleation.
\end{abstract} 

\begin{multicols}{2}
\narrowtext

Considerable interest is attached to determining the entire solitary 
wave  sequences of  solutions of the Gross-Pitaevskii (GP) model 
\cite{gp} because  they define possible states that can be excited 
in a Bose condensate. Jones and Roberts \cite{jr} determined the 
entire sequence
of solitary solutions   numerically for the GP model
$-2{\rm i} \psi_t = \nabla^2 \psi + (1 - |\psi|^2)\psi,$
where we use dimensionless variables
 such that the unit of length corresponds to the healing
length $a$, the speed of sound is $c=1/\sqrt{2}$,  and the density 
at infinity is $\rho_\infty=1$.
 The  solitary wave solutions satisfy two conditions: (1) the 
disturbance associated with the wave vanishes at large distances: 
$\psi\rightarrow 1$, $|{\bf x}|\rightarrow \infty$ and  (2) they 
preserve their form as they  propagate with a dimensionless velocity 
$U$, so that 
$
\psi(x,y,z,t)=\psi(x',y,z)$, $ x'=x-Ut,
$
 in three dimensions (3D)
and 
$
\psi(x,y,t)=\psi(x',y),
$
in two dimensions (2D), so that the solitary wave  solutions satisfy
\begin{equation}
2 {\rm i} U {\partial \psi \over \partial x'} 
= \nabla^2 \psi  + (1 - |\psi|^2)\psi.
\label{Ugp}
\end{equation}

 Jones and Roberts calculated the energy ${\cal E}$ and momentum ${\cal P}$ 
given by
\begin{eqnarray}
{\cal E}&=&{\frac{1}{2}}\int|\nabla \psi|^2\, dV + {\frac{1}{4}}\int 
(1 - |\psi|^2)^2\, dV \label{eJR}\\
{\cal P}      &=&{\frac{1}{2{\rm i}}}\int [(\psi^*-1)\nabla\psi-(\psi-1)
\nabla\psi^*]\, dV,
\label{pJR}
\end{eqnarray}
and determined  the location of the sequence on the ${\cal P}{\cal E}$-plane. 
  
In three dimensions they found two branches meeting at a cusp where 
${\cal P}$ and ${\cal E}$
assume their minimum values, ${\cal P}_{\rm min}$ and ${\cal E}_{\rm min}$. 
As ${\cal P} \to \infty$
on each branch, ${\cal E} \to \infty$. On the lower branch the
solutions are  asymptotic to  large vortex rings.

As ${\cal E}$ and ${\cal P}$ decrease from infinity along the
lower branch, the solutions begin to lose their similarity to
large vortex rings.
Eventually, for a momentum ${\cal P}_0$ slightly greater than ${\cal P}_{\rm min}$,
they lose their vorticity ($\psi$  loses its zero), and
thereafter the solitary solutions may better be described as
`rarefaction waves'. The upper branch consists entirely of these
and, as ${\cal P} \to \infty$ on this branch, the solutions asymptotically
approach the rational soliton solution of the Kadomtsev-Petviashvili Type I
(KPI) equation \cite{kp} and are unstable. 
In 2D the family of the solitary wave solutions are represented by two 
point vortices if $U\le 0.4$. As the velocity increases the wave loses 
its vorticity and becomes a rarefaction pulse. As $U\rightarrow 1/\sqrt{2}$  
both the energy, ${\cal E}$ and momentum ${\cal P}$ per unit length 
approach zero and the solutions asymptotically
approach the 2D rational soliton solution of KPI. 

Jones and Roberts \cite{jr} 
derived the KPI equation using an  asymptotic expansion in the  parameter 
$\epsilon^2\approx 2(1-\sqrt{2}U)$, which is small when  $U$ approaches the 
speed of sound. They sought  solutions of the form $\psi=f + {\rm i}g$, 
where $f=1+\epsilon^2 f_1 + \epsilon^4 f_2+\cdot\cdot\cdot$, 
$g=\epsilon g_1 + \epsilon^3 g_2+\cdot\cdot\cdot$, 
$U=1/\sqrt{2}+\epsilon^2 U_1+\cdot\cdot\cdot$.  The independent 
variables were stretched, so that $\xi=\epsilon x'$, $\eta=\epsilon^2 y$, 
and $\zeta=\epsilon^2 z$. By substituting these expressions into (\ref{Ugp}) 
and considering real and imaginary parts at the leading and  first orders 
in $\epsilon$, they determined that $g_1$ satisfies the KPI equation:
\begin{equation}
 {\frac{\partial^2g_1}{\partial\xi^2}} + \nabla^2_{\eta\zeta}g_1
-{\frac{\partial}{\partial \xi}}\biggl[{\frac{1}{2}}{\frac{\partial^3g_1}
{\partial\xi^3}}-{\frac{3}{\sqrt{2}}}\Bigl({\frac{\partial g_1}{\partial\xi}}
\Bigr)^2\biggr]=0,
\label{g1}
\end{equation}
and $f_1=\partial g_1/\sqrt{2}\partial \xi - g_1^2$, $U_1=-1/2\sqrt{2}$.
In 2D the equation (\ref{g1}) has a closed form solution \cite{man}, so that 
the asymptotic solution of the GP equation in the KPI limit is
\begin{equation}
\psi=1-{\rm i}{\frac{2\sqrt{2} x'}{{x'}^2 + \epsilon^2y^2 + 3/2\epsilon^2}}
-{\frac{2}{{x'}^2 + \epsilon^2y^2 + 3/2\epsilon^2}},
\label{kp}
\end{equation}
which we have written in the original variables.

It  was shown by Kuznetsov and Turytsin \cite{kt} that  the 2D KPI soliton 
is stable to 2D but  unstable to 3D perturbations. The linear stability 
analysis of 2D solitary solution of the GP equation subject to 
long wavelength infinitesimal perturbations was done by Kuznetsov 
and Rasmussen \cite{kuzras}. They demonstrated that all long wavelength 
antisymmetric modes are stable and all long wavelength symmetric 
modes are unstable. In particular they showed that the growth rate 
of symmetric perturbations, $\sigma$,  is given by 
$
\sigma^2=-{\cal E}k^2/(\partial{\cal P}/\partial U)>0,$ as the wavenumber 
$k\rightarrow 0.
$
The maximum growth rate of instability and the instability region of 2D 
solitary solutions were found in \cite{br2} by solving the linear stability
 problem. Through numerical integration of the GP equation it was shown  
that as perturbations grow to finite amplitude the vortex lines reconnect 
to produce a sequence of almost circular vortex rings. 
Senatorski and Infeld \cite{si}  numerically integrated  
the KPI equation to study
the fate of 2D  KPI solitons subject to 3D perturbations. They determined 
that 2D KPI solitons evolve into 3D KPI solitons which are also unstable.

 The goal of this Letter is to elucidate the fate of the 2D rarefaction 
pulse in the KPI limit of the GP model subject to 3D perturbations. We 
discovered that such solutions may evolve into vortex rings and this  
establishes a new mechanism of  vortex nucleation. We found that this 
mechanism can operate in different ways.
The intermediate states may involve  periodic solutions consisting of 
interacting 3D rarefaction pulses that belong to the lower 
branch of the Jones-Roberts cusp with ${\cal P}<{\cal P}_0$
%
%
%
%
%
%
%
%

We have performed  direct numerical simulations using the  numerical 
method described  in  \cite{br}. 
We solve the GP equation  in the reference frame moving
with the velocity $U_{\rm f}$ chosen in such a way that the main disturbance 
is kept within the computational box:
\begin{equation}
-2 {\rm i} \pd {\psi} t + 2 {\rm i} U_{\rm f} {\partial \psi \over \partial x} 
= \nabla^2\psi  + (1 - |\psi|^2)\psi.
\label{tUgp}
\end{equation}
In these computations we follow 
the evolution of the  asymptotic solution  (\ref{kp}) extended  along the 
$z-$axis and   moving in the $x-$direction.
The dimensions of  the computational box  are $D_x=60$, $D_y=60$, $D_z=180$. 
The $xy-$faces of the box are open, to allow sound waves
to escape; this is achieved numerically by applying the Raymond-Kuo
technique \cite{rk}. The $z=0$ and $z=D_z$ sides are reflective.

The soliton (\ref{kp}) was perturbed along the $z-$axis, so that at $t=0$
\begin{equation}
x\rightarrow x+0.1 \cos (kz).
\label{pert}
\end{equation}
We choose $k$ so that  $N$ periods of this perturbation fit exactly into 
the $D_z$ dimension of the box. There are two main parameters of the problem 
that determine the final outcome of the  instability: $\epsilon$, which 
determines the configuration, energy, and momentum of the initial field, 
and the wavelength of the perturbation $\ell=D_z/N$. It can be easily 
shown using (\ref{kp}) that the energy (\ref{eJR}) and the momentum 
(\ref{pJR}) of our initial field per wavelength of perturbation are 
given by
\begin{equation}
{\cal E}= {\cal P}/\sqrt{2}=8\pi\epsilon \ell/3.
\label{epl}
\end{equation}
First, we  consider the evolution of the KPI solitary solution subject to   
large wavelength perturbations $\ell=20, 30, 60$ and $\epsilon =0.5$.  Fig.\ 1
 illustrates the appearance of vortex rings through  contour plots of 
the cross-section of the solution in the $xz-$plane with $y=D_y/2$ for $\ell=20$. 
The last panel shows the isosurface $|\psi|^2=0.2$.
\begin{figure}
\caption{ \small The contour plot of the density field of the cross-section 
of  solutions of the GP equation. The time snapshots show the evolution 
of the KPI solution (\ref{kp}) of the GP equation in the $xz-$plane with $y=D_y/2$. 
In (\ref{kp}) we took $\epsilon=0.5$ and the wavelength of the initial 
perturbation (\ref{pert}) is $\ell=20$.  The last panel shows the isosurface 
$|\psi|^2=0.2$. The solutions starting with the third panel posess vorticity and evolve into  equally spaced vortex rings.}
\bigskip
\centering
\psfig{figure=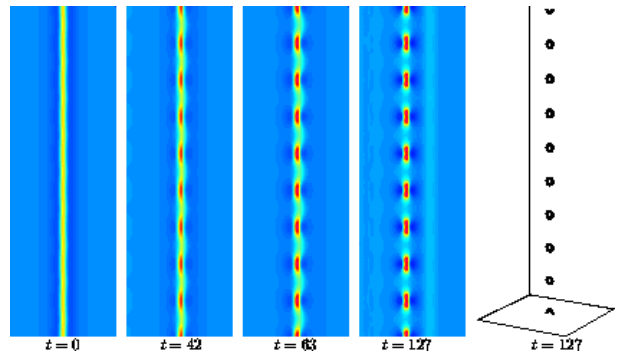,height=1.8in}
\end{figure}
%

According to the time snapshots of Fig.\ 1 the solution evolves directly 
into  a set of  vortex rings;   other axisymmetric 3D solutions including 
3D KPI solitons, are not involved.  Moreover, exactly one vortex ring is 
generated for each wavelength of the perturbation. These vortex rings  are 
distanced $\ell$ healing lengths apart and  have radii much smaller than 
$\ell$ (see Table 1). They therefore interact only weakly with each other. 
The energy and momentum (\ref{epl}) of one period of the perturbation is 
used to create one vortex ring, the extra energy and momentum escapes  and 
is carried away by  sound waves (phonons). The vortex rings are aligned 
and propagate together with the same velocity. This arrangement of vortex 
rings is itself unstable and cannot last forever.

Similar calculations were done for $\ell=30$ and $\ell=60$. The results 
are summarized in Table 1 which gives the energy and  momentum per wavelength 
of the perturbation  of  the   initial field and  the energy, momentum, 
velocity, and radius of the resulting vortex ring.

 {\bf Table 1.}
\medskip

\begin{tabular}{|c|c|c|c|c|c|c|}
\hline
$\ell$ & ${\cal E}_{{\rm init}}$ & ${\cal P}_{{\rm init}}$  &
$ {\cal E}_{{\rm ring}}$ & ${\cal P}_{{\rm ring}}$ & $U_{\rm ring}$& 
$R_{\rm ring}$  \\ \hline
 60 & 251  & 355  & 99 & 162 & 0.45 & 2.7  \\
 30 & 126 & 178  & 86 & 132 & 0.49 & 2.35\\
 20 & 84 & 120    & 71 & 102 & 0.53 & 1.9 \\ \hline
\end{tabular} 
\bigskip
%
%
%
%
%
%
%
%
%
%

Next we  explore the evolution of the KPI limit solitary waves of the GP 
model subject to small wavelength  perturbations.
The effect of the decrease in the wavelength of the perturbation is 
twofold:
1) the energy and the momentum (\ref{epl}) per wavelength  of the initial 
field  is decreased, therefore leaving less energy available for creating 
a new entity and 2) these entities are in close proximity to  each other 
so they strongly interact. 
 Fig.\ 2 plots the cross-sections of the solutions for $\epsilon=0.5$ and 
$\ell=60, 30, 15, 7.5, 3.75$. The first two panels (Fig.\ 2a and 2b) 
illustrate the vortex nucleation discussed earlier. To the best of our 
knowledge the periodic solutions shown on Fig. 2c-e are unknown in the 
literature on the  GP model or the  nonlinear Schr\"odinger equation. 
The interesting feature of these solutions is that they lack a 
vorticity and have small energy and momentum per period that tend to 
zero as $\ell\rightarrow 0$. These solutions can be understood as periodic 
pulse trains composed of the rarefaction pulses positioned on the lower 
branch of the Jones-Roberts sequence with ${\cal P}<{\cal P}_0$. 
The interaction between 
adjacent pulses reduces the total energy  per period. The analysis of 
these and other properties of  periodic pulse trains composed from the 
solitary waves of nonintegrable evolution equations can be found in \cite{bh}. 
\begin{figure}
\caption{ \small The contour plot of the density field of the 
cross-section of  solutions of the GP equation. The time 
snapshots show the different stages in the evolution of the KPI-limit 
solution (\ref{kp}) of the GP equation in the $xz-$plane with $y=D_y/2$. 
In (\ref{kp}) we took $\epsilon=0.5$ and the wavelengths of the 
initial perturbation (\ref{pert}) were  $\ell=60$ (a), $\ell= 30$ (b),$\ell=  15$ (c),$\ell=  7.5$ (d), and $\ell= 3.75$ (e). The contour 
plots are shown for $|\psi|^2$ at $t=149$ for (a) 
and $t=127$ for the rest of panels.}
\bigskip
\centering
\psfig{figure=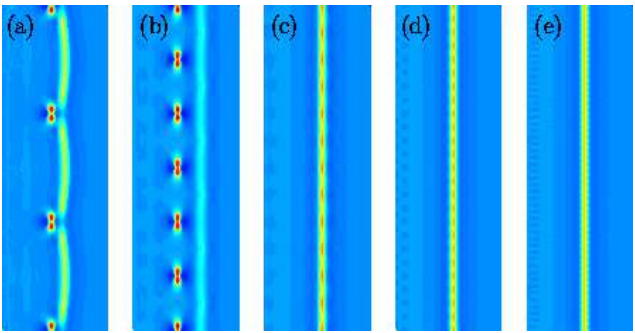,height=1.7in}
\end{figure}
Other findings are summarized in Table 2 which gives the values of the 
energy and  momentum of the initial field and the resulting periodic 
rarefaction solution.

 {\bf Table 2.}
\medskip

\begin{tabular}{|c|c|c|c|c|}
\hline
$\ell$ & ${\cal E}_{{\rm init}}$ & ${\cal P}_{{\rm init}}$ &
$ {\cal E}_{{\rm per}}$ & ${\cal P}_{{\rm per}}$  \\ \hline
 15 & 62 & 90 & 55 & 78   \\
 7.5 &31.5 &46.2 &30.7 &44 \\
  3.75 &15.5 &23 &15 &21   \\\hline
\end{tabular} 
\bigskip

These periodic  solutions can execute   standing wave  oscillations  of 
 decreasing amplitude and period $\ell/2$. 
Similarly to the periodic solutions made of  aligned vortex rings,  
the periodic rarefaction solutions become  unstable and we followed the 
development of this instability. Fig.\ 3a shows the contour plots of the 
density for a  cross-section of the field  which evolves from  the periodic 
rarefaction solution with $\ell=15$ to produce  total of four rings. Therefore, to create each ring the energy 
and momentum of several periods  of the rarefaction solution were used. 
The wavelength of the secondary instability that destroyed the periodic 
solution is approximately   $57$. Similar calculations were done for the even 
shorter perturbation wavelength $\ell=7.5$; see Fig.\ 3b. 
The wavelength of the secondary instability is approximately  $29$ 
resulting in the appearance of six rings. The reason for the apparent 
nonuniformity of the nucleated rings is that $D_z$ is not an exact 
multiple of these wavelengths.
\begin{figure}
\caption{ \small The contour plot of the density field of the 
cross-section of  solutions of the GP equation. The time 
snapshots show the different stages in the evolution of the KPI-limit 
solution (\ref{kp}) of the GP equation with $\epsilon=0.5$  
in the $xz-$plane with $y=D_y/2$. The wavelengths of the initial 
perturbation (\ref{pert}) are  $\ell=15$ (a) and $\ell=7.5$ (b).}
\bigskip
\centering
\psfig{figure=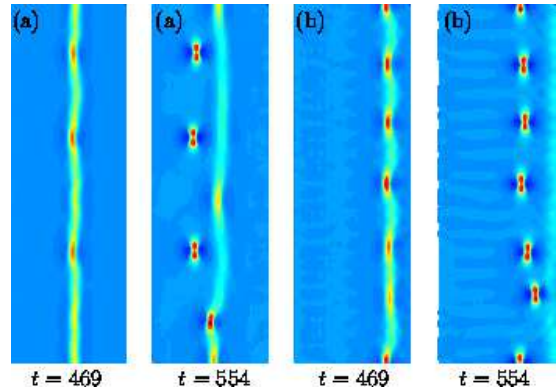,height=2in}
\end{figure}
%
%
%
%
%
%
%
%
%

Finally,  we consider the evolution of the small energy and momentum 
KPI limit solitary waves. In these computations we use $\epsilon=0.3$ 
and $\ell=60,30, 20, 15$.  For $\ell=60$ the KPI solution follows the 
scenario of vortex nucleation and directly  evolves into 3 vortex rings 
of small radii.
For smaller wavelengths $\ell =30, 20, 15$ the KPI solution initially evolves 
into oscillating periodic rarefaction pulses of decreasing $y-$extent. The 
energy and momentum per  period are apparently insufficient to allow 
them to evolve into rings and the necessary energy cannot be reduced 
through interactions when the putative solutions are separated by such 
large distances. These solutions break down into sound waves that carry 
off  all   energy and momentum; see Fig.\ 4.
\begin{figure}
\caption{ \small Isosurfaces of the density field of  solutions 
of the GP equation. The time 
snapshots show the different stages in the evolution of the KPI-limit 
solution (\ref{kp}) of the GP equation with $\epsilon=0.3$. The wavelengths of the initial 
perturbation  (\ref{pert}) is  $\ell=15$. The minimum density  increases with time and approaches unity as the solution breaks down into sound waves.}
\centering
\psfig{figure=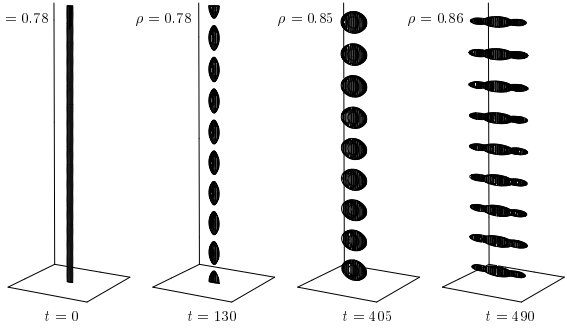,height=1.8in}
\end{figure}
%
%
%
%
%
%
%

In summary, we studied the  instability of  the 2D KPI limit solitary 
wave solution  in the GP equation. The evolution of several types of 
solutions are considered. We were able to identify three different 
regimes of transition depending on the initial energy and momentum of the 
KPI solution and on the wavelength of the initial perturbation. For large 
wavelengths, the initial solution immediately evolves into a periodic 
solution consisting of  small equally spaced vortex rings  with a 
period equal to the  period of the initial perturbation. For shorter 
wavelengths the KPI solution first evolves into a periodic solution 
consisting of 3D interacting rarefaction pulses that later break up 
into vortex rings under the influence of a secondary instability of a 
different wavelength. Finally, if the energy of the KPI solution is 
small  the solution can break into sound waves 
 after forming an  oscillating periodic 
rarefaction pulse. 

Fig.\ 5 summarizes all calculations performed and the relationships of 
the different regimes studied. The initial states considered  are 
represented  by dots on the ${\cal P}{\cal E}-$ plane, where ${\cal E}$ and ${\cal P}$ 
are defined per wavelength of the perturbation. 
We  plot the cusp determined by Jones and Roberts \cite{jr} for the 
family of the vortex rings and rarefaction pulses. 
The arrows show the way the 
initial state evolve.
\begin{figure}
\caption{ \small  Summary of numerical integration of (\ref{tUgp})  starting 
with the initial condition (\ref{kp})  with  $\epsilon=0.5$  and  
$\epsilon=0.3$. The cusp  corresponds to the 3D solitary wave solutions.       Dashing of the   upper branch 
indicates that this branch is unstable. 
(${\cal P}_0, {\cal E}_0$) marks the point where the vorticity 
disappears and point (${\cal P}_{\rm min}, {\cal E}_{\rm min}$) gives 
the position of the lowest momentum-energy state of the 3D solitary 
solution (see discussion in the text).  The line from the origin to $({\cal P}_{\rm min},{\cal E}_{\rm min})$ corresponds to the family of  periodic 
rarefaction pulses.  Dots indicate the position of the initial states, 
the wavelength of the perturbation $\ell$ is given next to each initial  
state; 
arrows show the evolution of these solutions, and the crosses correspond 
to the final state before the onset of the secondary instability.
}
\centering
\psfig{figure=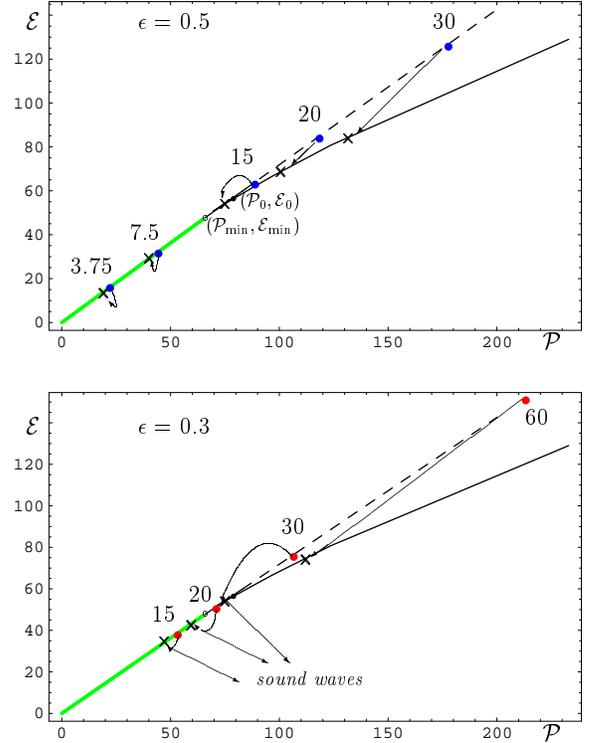,height=4in}
\end{figure}
%

This work was supported by the NSF grants DMS-9803480 and DMS-0104288. 
The author is very grateful to Professor Paul Roberts for many useful 
discussions about this work.

\end{multicols}
\end{document}